\documentclass{article}
\usepackage{PRIMEarxiv}
\usepackage[utf8]{inputenc} 
\usepackage[T1]{fontenc}    
\usepackage{hyperref}       
\usepackage{url}            
\usepackage{booktabs}       
\usepackage{amsfonts}       
\usepackage{nicefrac}       
\usepackage{microtype}      
\usepackage{lipsum}
\usepackage{placeins}
\usepackage{float}
\usepackage{fancyhdr}       
\usepackage{graphicx}       
\graphicspath{{media/}}     

\usepackage{listings} 

\usepackage{amsmath}

\title{Adapting Large Language Models to Mitigate Skin Tone Biases in Clinical Dermatology Tasks: A Mixed-Methods Study
}
\author{
  Kiran Nijjer\textsuperscript{1,2}, Ryan Bui\textsuperscript{1,2}, Derek Jiu\textsuperscript{1,2}, Adnan Ahmed\textsuperscript{1,2}, Peter Wang\textsuperscript{1,2}, Kevin Zhu\textsuperscript{2}, Lilly Zhu\textsuperscript{2} \\
  \small Stanford University, Stanford, United States, Algoverse AI Research, Palo Alto, United States
}
\date{April 2025}

\begin{document} 
\maketitle 
\begin{abstract}

SkinGPT-4, a large vision-language model, leverages annotated skin disease images to augment clinical workflows in underserved communities. However, its training dataset predominantly represents lighter skin tones, limiting diagnostic accuracy for darker tones. Here, we evaluated performance biases in SkinGPT-4 across skin tones on common skin diseases, including eczema, allergic-contact dermatitis, and psoriasis using the open-sourced SCIN dataset. We leveraged the SkinGPT-4 backbone to develop finetuned models for custom skin disease classification tasks and explored bias mitigation strategies. Clinical evaluation by board-certified dermatologists on six relevant skin diseases from 300 SCIN cases assessed images for diagnostic accuracy, informativity, physician utility, and patient utility. Model fairness metrics, including demographic parity and equalized odds, were calculated across skin tones. SkinGPT-4 achieved an average demographic parity of 0.10 across Fitzpatrick types, with notable differences of 0.10–0.15 between lightest and darkest tones across evaluation metrics. Model hallucinations in artifacts and anatomy occurred at a rate of 17.8. Our customized models achieved average F1, precision, and AUROC of 0.75, 0.78, and 0.78 across visually similar disease pairs. Fairness analysis showed an average demographic parity of 0.75, with a maximum disparity of 0.21 across skin tones. The best model achieved parity scores of 0.83, 0.83, 0.76, 0.89, 0.90, and 0.90 for Fitzpatrick I–VI, indicating robust fairness. Large language models such as SkinGPT-4 showed weaker performance on darker tones. Model biases exist across evaluation criteria, and hallucinations may affect diagnostic efficacy. These findings demonstrate the efficacy of training accurate, fair models using existing backbones for custom skin disease classification. [1]
\end{abstract}

\section{Introduction}

The integration of generative pre-trained transformers (GPTs) is reshaping dermatological practice through its ability to interpret and diagnose dermoscopy images. GPTs have significantly advanced image classification by accurately identifying various skin diseases and lesions with performance equal or superior to that of professional dermatologists \cite{BJD2020Diversity} \cite{SDR2019Outperform} \cite{SDR2019Equal} . As research in artificial intelligence continues to expand, its exploration within dermatology has led to advancements in diagnostic imaging, disease classification, and personalized treatment planning \cite{Nature2017SC} \cite{EJCancer2019Meta}. With the heavy shortage of dermatologists \cite{Frontiers2023Overview} \cite{JAMAderm2018FewDerms} \cite{JAAD2008Access} and lack of access to their services in many areas throughout the US \cite{Frontiers2023Overview} \cite{TMJ2018LimitedDerm} \cite{JAAD2006Access}, GPTs have quickly gained traction as a complementary diagnostic tool to treat patients. Continually improving its ability to analyze patterns through training \cite{BMJ2023AI}, GPTs present a promising avenue for independent artificial intelligence (AI) driven healthcare models for dermatology diagnosis and treatment. However, the significance of equitable performance across diverse populations is a growing concern as models trained on imbalanced datasets have the risk of reinforcing existing healthcare disparities \cite{obermeyer2019dissecting} \cite{char2018implementing}. 

SkinGPT-4 is the first large language model (LLM) able to provide interactive, dermatological text diagnosis of skin images \cite{Zhou2024SkinGPT4}. SkinGPT-4 leverages MiniGPT-4, a vision-based large language model (LLM), fine tuned with a large dataset of over 50,000 skin disease images with corresponding clinical annotations and is trained to include trusted medical knowledge of skin diseases. Users can input their own photos to SkinGPT-4 for diagnosis, and SkinGPT-4 will output text evaluations along with potential treatment suggestions. The system’s local deployment ensures user privacy and makes it easily accessible, even in underserved regions with scarce dermatological resources. SkinGPT-4 streamlines skin disease diagnosis and education by delivering rapid and accessible care to everyone and anywhere.

Despite its strong potential to bridge gaps in healthcare delivery, Skin-GPT faces challenges related to demographic bias, most notably its handling of skin tone groups. Lighter skin tones dominate the make-up of existing dermatological datasets, while darker skin tones remain significantly underrepresented \cite{tadesse2023skin} \cite{louie2018representations}. 

For example, up to 80 percent of images in the Fitzpatrick 17k and SkinCAP datasets, both widely used for skin condition research, depict skin diseases on white skin, while all other ethnic groups—including Asian, African, Hispanic, South Asian, Pacific Islander, and others— collectively make up only around 20 percent of the images \cite{Fitzpatrick17kArXiv} \cite{Fitzpatrick17kGitHub} \cite{Fitzpatrick1988Validity} \cite{Fitzpatrick17kArXiv}. Because skin diseases look different across skin tones \cite{tadesse2023skin}, these color representation inequalities within datasets create prediction bias, which leads to diagnostic inaccuracies and delayed treatment for affected populations \cite{adamson2018machine} \cite{chen2024unmasking}.

Ensuring high diagnostic accuracy in Skin-GPT is crucial for diagnosing severe or life-threatening conditions, protecting patient well-being, and facilitating timely, appropriate medical treatment. To address these inequalities, a strong emphasis should be placed on having frameworks that incorporate equitable demographic representation into model training in order to ensure these AI tools perform fairly across different skin tones \cite{seyyed-kalantari2021underdiagnosis} \cite{ACL2017PRR}.

This study systematically evaluates the demographic bias of Skin-GPT by analyzing its performance across different skin types. Using a weighted loss function, we will stratify datasets by skin tone to target any disparities in diagnostic accuracy across these skin tones. In addition, our study utilizes a multi-layer perceptron (MLP) head and bias-aware training mechanisms to mitigate skin tone bias and enhance SkinGPT-4 performance across all skin tones, informing the development of more equitable and effective AI models in dermatology. 
\begin{figure}[H]
    \centering
    \includegraphics[width=0.9\linewidth]{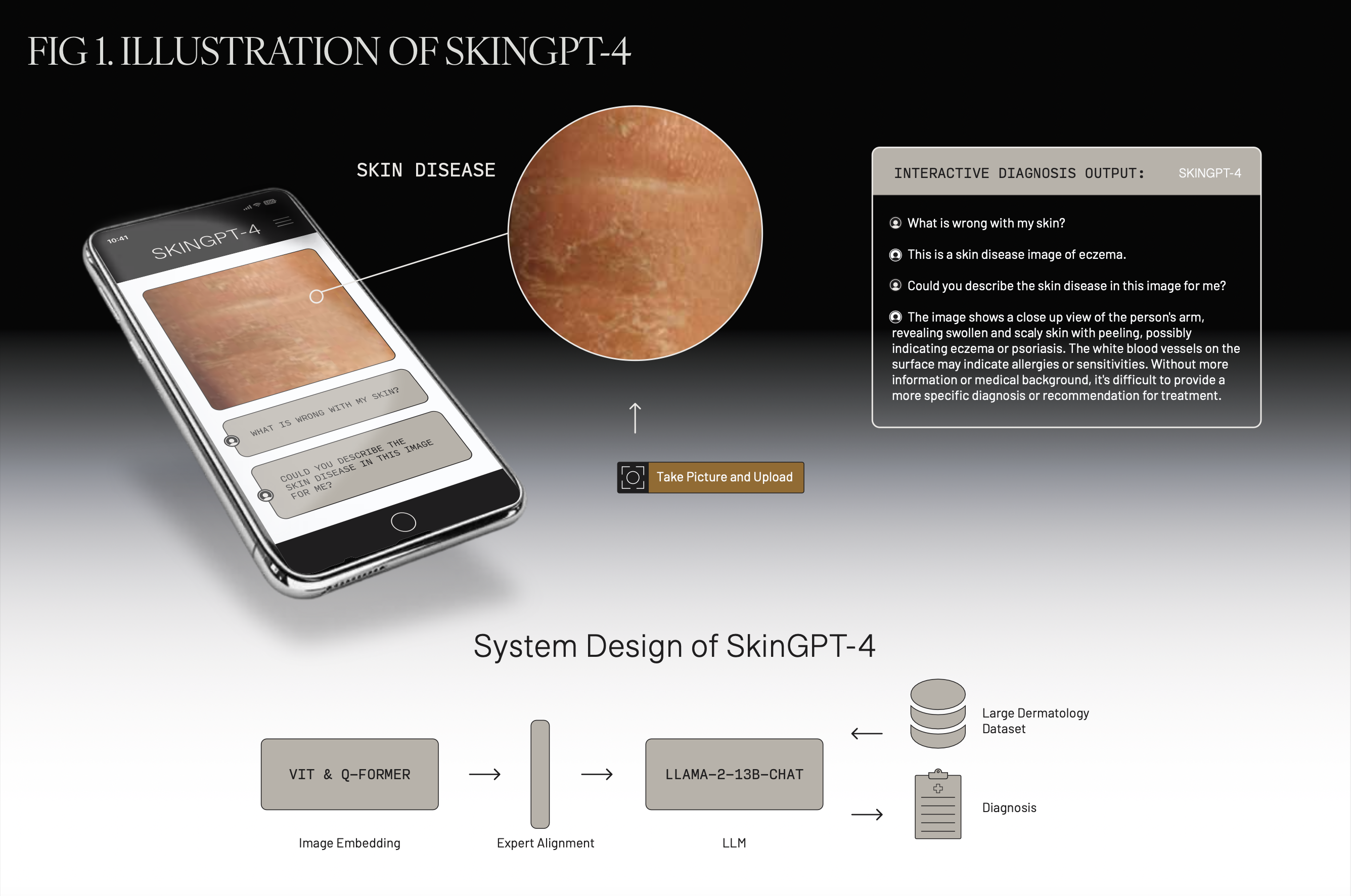}
    \caption{Illustration of the original SkinGPT-4 model architecture. An example initiated by a user and the LLM agent are depicted.}
    \label{fig:skinmodel}
\end{figure}
\section{Related Work}
In recent years, there’s been growing research on the use of artificial intelligence in dermatology, especially with tools like convolutional neural networks (CNNs) and transformer-based models. These technologies are being used to help classify skin lesions and diagnose various skin conditions from images. Previous research has shown that these models can match the accuracy of dermatologists when identifying skin cancers and other dermatological diseases by training on large datasets of medical images \cite{Nature2017SC} \cite{Han2018}.

Similarly, there has been increasing concern about bias in AI systems used for medical diagnosis. When training data lacks diversity such as sufficient images for darker skin tones, then these AI models are likely to underperform for patients of color \cite{adamson2018machine}. This may result in these models misidentifying or overlooking conditions on darker skin, which can lead to delayed or incorrect treatment. These issues point to the urgent need for more inclusive training data.

To tackle these inequalities, researchers have looked into ways to measure and reduce bias in AI models. Approaches like re-sampling datasets and adjusting weights during training have been suggested to help models perform more fairly across different demographic groups \cite{seyyed-kalantari2021underdiagnosis} \cite{Holstein2019} \cite{Mehrabi2021}. To assess fairness, tools such as equal opportunity difference, and accuracy parity are now commonly used in evaluating medical AI systems. The importance of continuously auditing AI systems to identify and mitigate biases is crucial, which will ultimately support fair and effective clinical diagnosis for all patient groups. 

There is a critical need for addressing systemic inequities in AI development to avoid perpetuating or worsening healthcare disparities. These perspectives emphasize that beyond technical performance, fairness and equity must be core components of AI system design. Our work expands on these prior studies by introducing SkinGPT-4, a vision language model (VLM)-based diagnostic tool fine-tuned on dermatological datasets, and clinically evaluating its diagnostic performance across different skin tones. While earlier research has largely focused on CNN-based models, few have explored how vision-language models like the ones used in SkinGPT-4 can inherit or amplify demographic biases. 

Moreover, our study uniquely incorporates a weighted loss function stratified by skin tone, a multi-layer perceptron (MLP) head and bias-aware training pipeline to actively mitigate bias in the model. By combining fairness evaluation with bias mitigation strategies in a vision-language model context, this study contributes a novel framework for equitable AI development in dermatology.

\section{Methodology}
A clinical evaluation of SkinGPT-4 was performed by a board-certified dermatologist on six clinically relevant skin diseases, including eczema, allergic contact dermatitis, and psoriasis, from 300 cases in the open-sourced SCIN dataset. The diseases chosen for this evaluation were selected for their equal representation across different skin tones in the SCIN dataset, ensuring a balanced assessment of SkinGPT-4’s performance across a variety of conditions.

Images from the dataset were assessed on multiple criteria: diagnostic accuracy, informativity, physician utility, and patient utility. These criteria were defined as follows: diagnostic accuracy measured the model's ability to correctly identify the disease, informativity assessed the usefulness of the model's output for guiding clinical decision-making, physician utility referred to how effectively the model’s predictions aided clinicians in their diagnostic process, and patient utility measured the clarity and relevance of the model’s output from a patient’s perspective.

To evaluate model fairness, demographic parity and equalized odds metrics were calculated for each skin tone, using the Fitzpatrick scale identifications provided in the dataset. These fairness metrics were computed for each of the evaluated conditions to assess whether there were disparities in the model’s performance across different skin tones. Additionally, the evaluation accounted for model hallucinations—instances where the model generated incorrect predictions regarding physical artifacts or body anatomy—by tracking the rate of such occurrences across the tested conditions. 

For the methods section of our study, we selected 600 cases from the SCIN dataset with balanced representation across Fitzpatrick skin tones 1–6 to ensure equitable evaluation. The original SkinGPT-4 model architecture incorporates Vision Transformer (VIT) and Q-Former modules for generating image embeddings. By passing through an alignment layer, the image embeddings are parsed and aligned to generate a Llama-2-13b text output. 

Our fine-tuned model improves on this architecture by freezing the vision transformer layers and optimizing hyperparameters to suit the dataset characteristics. A custom multi-layer perceptron (MLP) classification head was appended to predict one of six disease labels. To evaluate bias and fairness, we implemented metrics from Fairlearn, assessing demographic parity, true positive rate (TPR), false positive rate (FPR), bias difference, and bias ratio across demographic groups. Equalized odds and demographic parity were also examined to further assess fairness. 

Model performance was evaluated using standard metrics including accuracy, precision, recall, and area under the receiver operating characteristic curve (AUROC), with particular attention to generalizability across all six Fitzpatrick skin tones. To improve model results, we implemented an oversampling algorithm to emphasize underrepresented disease classes.The model was re-trained with these mitigation strategies and reassessed for fairness improvements. 

\clearpage

\begin{figure}[H]
    \centering
    \includegraphics[width=0.9\linewidth]{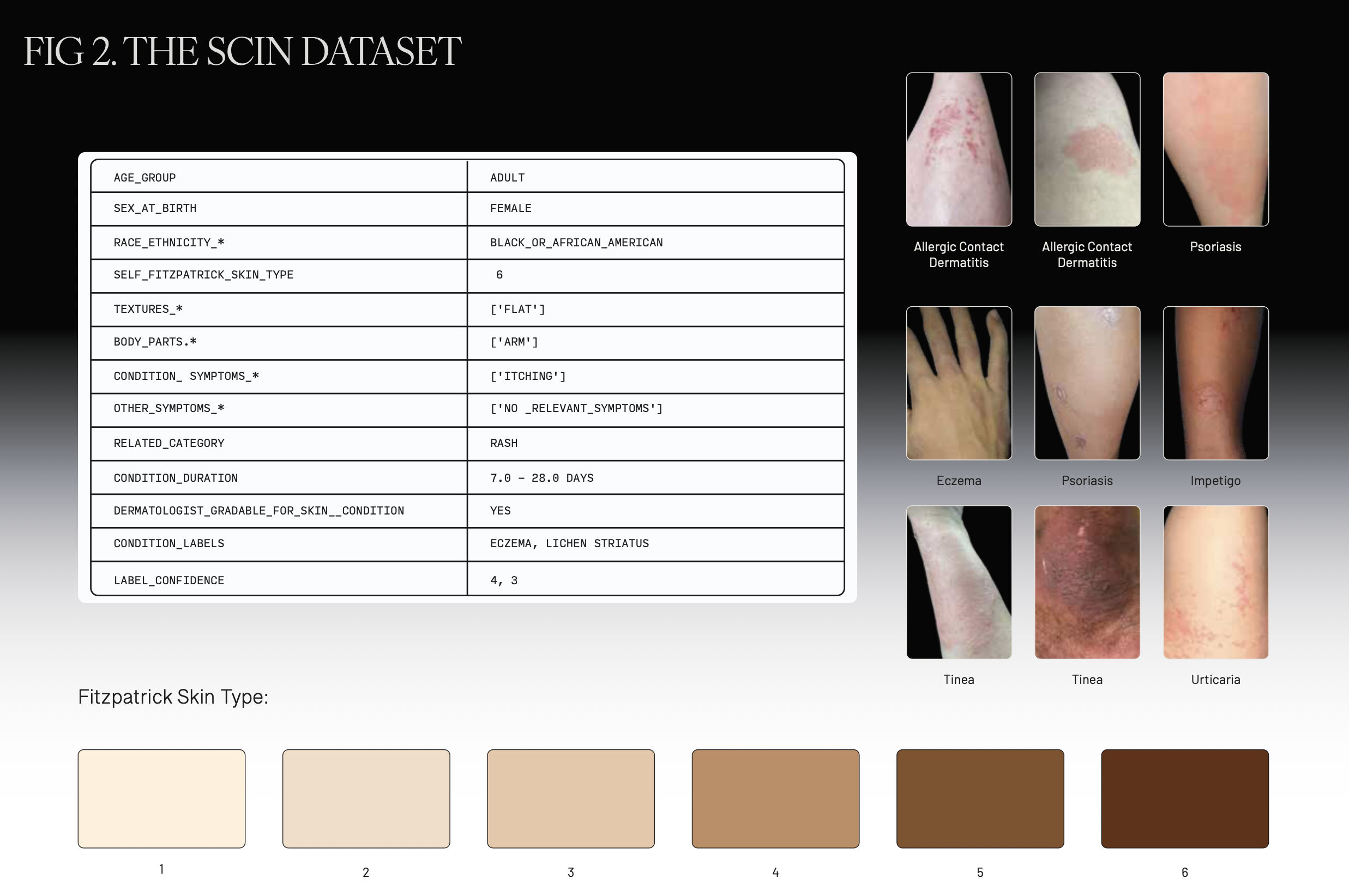}
    \caption{Illustration of the SCIN Dataset, including its accessible fields and sample images, used in the evaluation of SkinGPT-4.}
    \label{fig:enter-label}
\end{figure}

\section{Results}
SkinGPT-4 demonstrated a moderate level of demographic parity across all skin tones on the Fitzpatrick scale, achieving an average demographic parity of 0.10. Notable disparities were observed in the performance metrics for different skin tones. Specifically, demographic parity differences of 0.10, 0.10, 0.11, and 0.15 were observed between the lightest and darkest skin tones relative to diagnostic accuracy, informativity, physician utility, and patient utility, respectively. Model performance metrics for SkinGPT-4 are summarized in Figure 3. 

The diagnostic accuracy of the finetuned model was 0.8095, with an area under the precision-recall curve (AUPRC) of 0.8613 and an area under the receiver operating characteristic curve (AUROC) of 0.9082. The F1 score was 0.8000, with precision and recall both achieving scores of 0.8083 and 0.8095, respectively.

In terms of fairness, as illustrated in Figure 4, SkinGPT-4 achieved an average parity of 0.531, with the largest parity difference observed at 0.500. These values reflect the model's tendency to show varying levels of performance across different skin tones.

Despite its promising performance, SkinGPT-4 was prone to model hallucinations, particularly in the context of physical artifacts and body anatomy. These occurred at a rate of 17.8 percent, indicating that the model occasionally misinterpreted or failed to correctly identify features in the images.

\clearpage
\begin{figure}[H]
    \centering
    \includegraphics[width=0.9\linewidth]{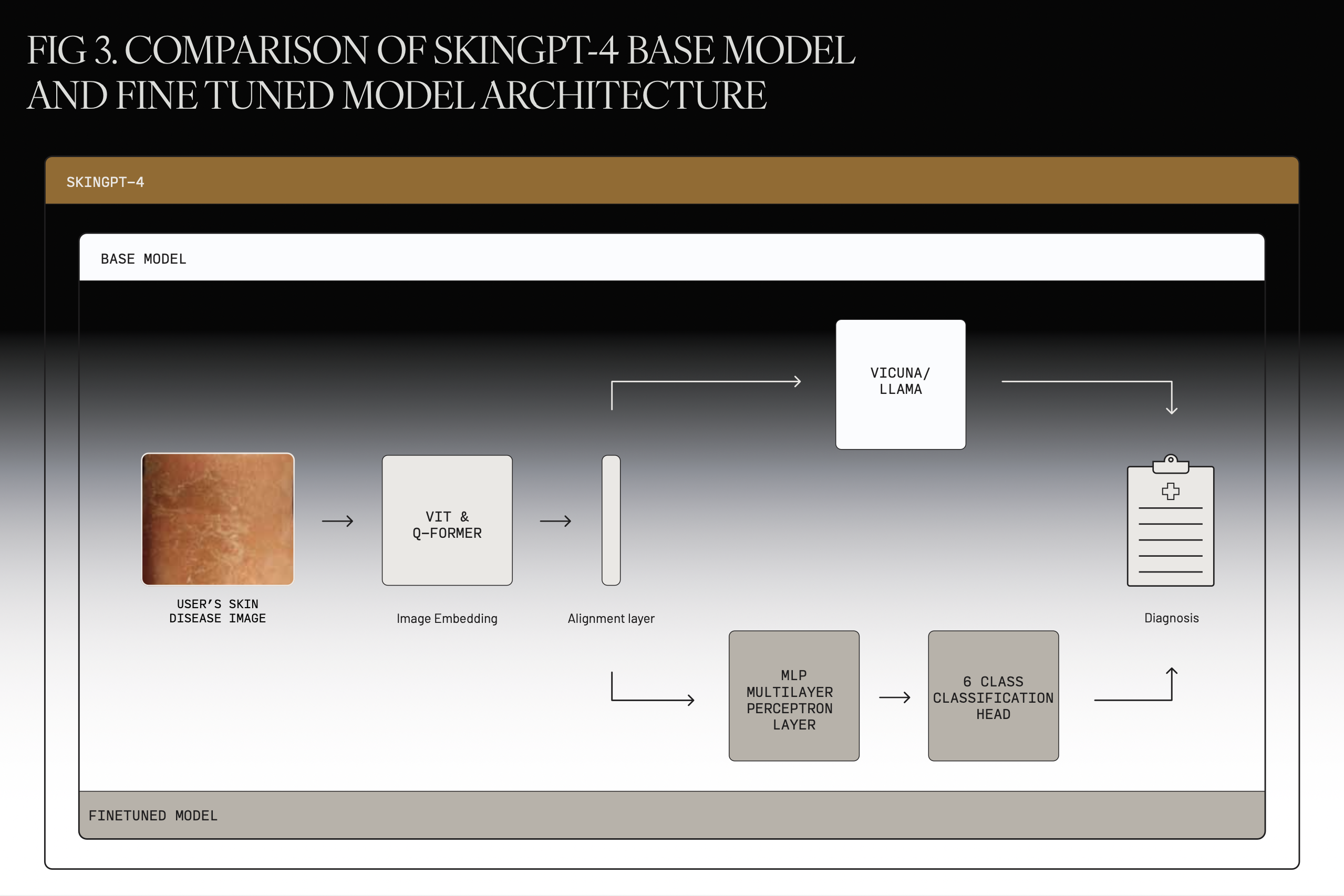}
    \caption{An illustration of the architectural differences between the base and fine-tuned model of SkinGPT-4.}
    \label{fig:enter-label}
\end{figure}

\section{Discussion}

Feedback from the dermatologist involved in the evaluation of SkinGPT-4, provided valuable insights into the model's clinical utility. The board-certified dermatologist highlighted instances where SkinGPT-4 incorrectly identified conditions, particularly when it suggested an infection that was not present, failed to recognize a rash, or overly generalized diagnoses (such as labeling a rash as either an infection or allergy). These misclassifications led to concerns regarding the model’s diagnostic accuracy in specific cases.

Additionally, the dermatologist emphasized that she trusted the model’s diagnosis when it accurately listed a validated diagnosis within its differential. However, there were cases where SkinGPT-4 correctly identified the condition but provided descriptions that did not align with the clinical image or diagnosis. For instance, in one case, the model described an image of a large, pink bruise on the knee as a result of a fall or injury. This description, however, did not match the actual clinical presentation, creating discrepancies between the model's output and the observed condition. Such inconsistencies raise concerns about the model’s ability to provide clinically relevant and accurate descriptions in addition to diagnoses. These findings highlight areas for improvement in both the model's diagnostic accuracy and its descriptive capabilities.

This study has several limitations that should be considered when interpreting the results. First, the evaluation of SkinGPT-4 was conducted by a limited number of dermatologists—only three board-certified dermatologists participated in the evaluation process. While these experts provided valuable insights, the small sample size of evaluators may limit the generalizability of the findings.

Moreover, the selection of diseases for this study was based on specific criteria, including clinical relevance and similarity in clinical presentation across different skin tones. While these factors were considered carefully, they may introduce limitations in the diversity of conditions represented in the dataset. The diseases chosen may not fully encompass the broad spectrum of dermatological conditions encountered in clinical practice, which could affect the model’s performance on diseases outside of this selection.

\section{Conclusion}

The growing integration of AI in dermatology has highlighted both its potential and its limitations, particularly concerning demographic biases. While prior studies, including those by Esteva et al. \cite{Nature2017SC} and Han et al. \cite{han2018classification}, have demonstrated AI’s capabilities in diagnosing skin conditions, they have largely overlooked the critical issue of demographic fairness. Research by Adamson and Smith \cite{adamson2018machine} has illuminated the risks of training AI models on non-representative datasets, which can disproportionately affect patients with darker skin tones. These limitations emphasize the urgent need for AI systems that are not only accurate, but equitable across diverse populations. 

By introducing a multilayer perceptron (MLP) head we established a model that performs with greater equity across diverse demographic groups and makes it better at classifying diseases. This contributes to the growing body of work on developing AI systems that are not only accurate but also fair. In addition, our study provides a valuable benchmark for evaluating dermatological bias in AI, offering a framework that can guide future research in the field.

As AI continues to permeate healthcare, it is essential that these technologies are equitable and provide accurate care to all individuals, regardless of demographic characteristics. By highlighting the need for continuous auditing, bias mitigation, and fairness evaluation, our work emphasizes the role of AI in advancing healthcare equity. 

Through this research, we are fostering a more inclusive approach to AI development, ensuring that AI models in dermatology and beyond can serve the needs of all patients, thereby driving forward a future of fair, unbiased, and accessible healthcare for all.

\subsection{Appendix 1}
The SCIN dataset used in this study consists of annotated skin disease images covering a wide variety of dermatological conditions. The dataset contains over 10,000 images of dermatology conditions, encompassing skin, nail and hair conditions. These images were crowdsourced with informed consent from US users and were annotated by dermatologists, providing reliable clinical annotations for each case. We assessed 300 cases of six clinically relevant skin diseases: eczema, allergic-contact dermatitis, psoriasis, tinea, urticaria, and others, ensuring a representative sample for evaluation. 
The SCIN dataset includes images spanning the Fitzpatrick scale (I-VI). However, the dataset is predominantly comprised of lighter skin tones, with darker skin tones (Fitzpatrick types V and VI) being underrepresented \cite{SCIN2023}. The skin tone distribution in SCIN is as follows--
Type I: 7.5\%
Type II: 21.7\%
Type III: 26.4\% 
Type IV: 17.1\%
Type V: 8.5\%
Type VI: 5.7\%

\subsection{Appendix 2}
SkinGPT-4 is based on MiniGPT-4, a vision-based large language model (LLM). The model combines a vision encoder that processes skin disease images with a text decoder that generates diagnostic information. A learnable MLP head was added to the encoder for improved skin disease classification. The following layers were used in the architecture:
Vision encoder: Pretrained MiniGPT-4 vision transformer
Text decoder: Pretrained language model layer
MLP head: Added to the encoder for classification task. 
Training Details
Training Dataset: SCIN dataset
Epochs: Maximum at 200
Batch Size: Training batch size = 16
Learning Rate: 0.0005
Optimizer: Adam
Loss Function: Cross Entropy 
Regularization: N/A

\subsection{Appendix 3}
 Demographic parity measures the equality of outcomes (i.e., diagnostic accuracy) across different groups. It is defined as the difference in the percentage of correct diagnoses between groups. 
Equalized odds is a fairness metric that ensures equal false positive and false negative rates across groups. It is defined as:

\begin{align*}
\text{True Positive Rate (TPR):} \quad 
& P(\hat{Y} = 1 \mid Y = 1, \text{group}) = P(\hat{Y} = 1 \mid Y = 1, \text{other group}) \\
\text{False Positive Rate (FPR):} \quad 
& P(\hat{Y} = 1 \mid Y = 0, \text{group}) = P(\hat{Y} = 1 \mid Y = 0, \text{other group})
\end{align*}

The F1 score is defined as:
\[
F_1 = \frac{2 \cdot \text{Precision} \cdot \text{Recall}}{\text{Precision} + \text{Recall}}
\]

 \subsection{Appendix 4}
 Model’s performance metrics across skin tone groups for each skin disease
  \begin{figure}[H]
    \centering
    \includegraphics[width=0.8\linewidth]{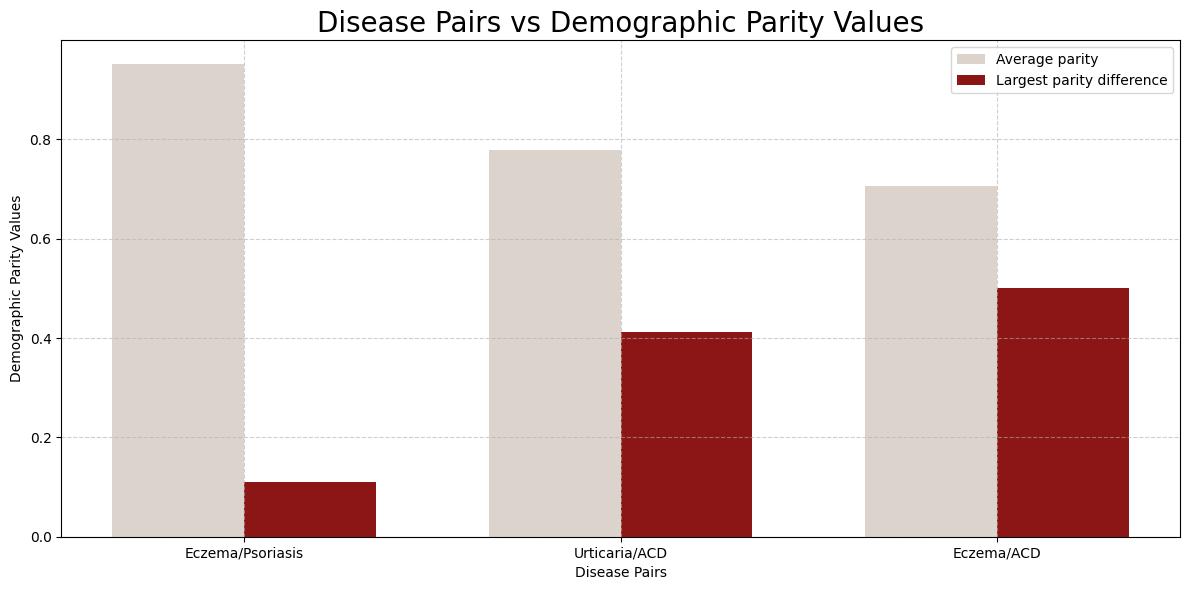}
    \caption{Relationship between disease pairs and their demographic parity values, illustrating disparities across conditions.}
    \label{fig:enter-label}
\end{figure}
\begin{figure}[H]
    \centering
    \includegraphics[width=0.8\linewidth]{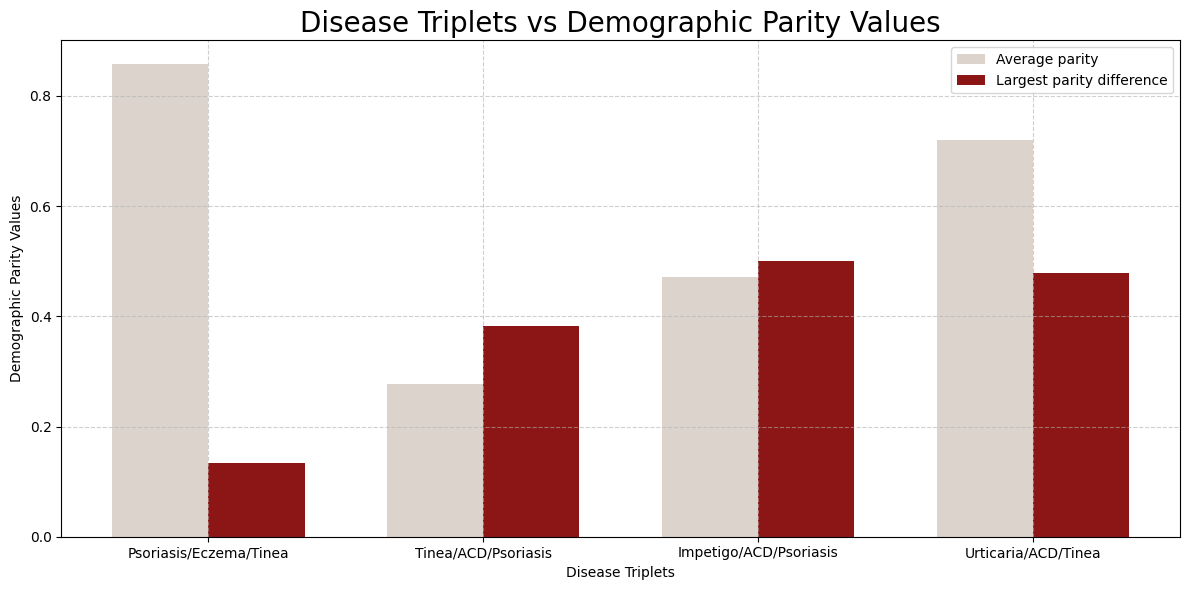}
    \caption{Relationship between disease triplets and their corresponding demographic parity values.}
    \label{fig:enter-label}
\end{figure}
\begin{figure}[H]
    \centering
    \includegraphics[width=0.8\linewidth]{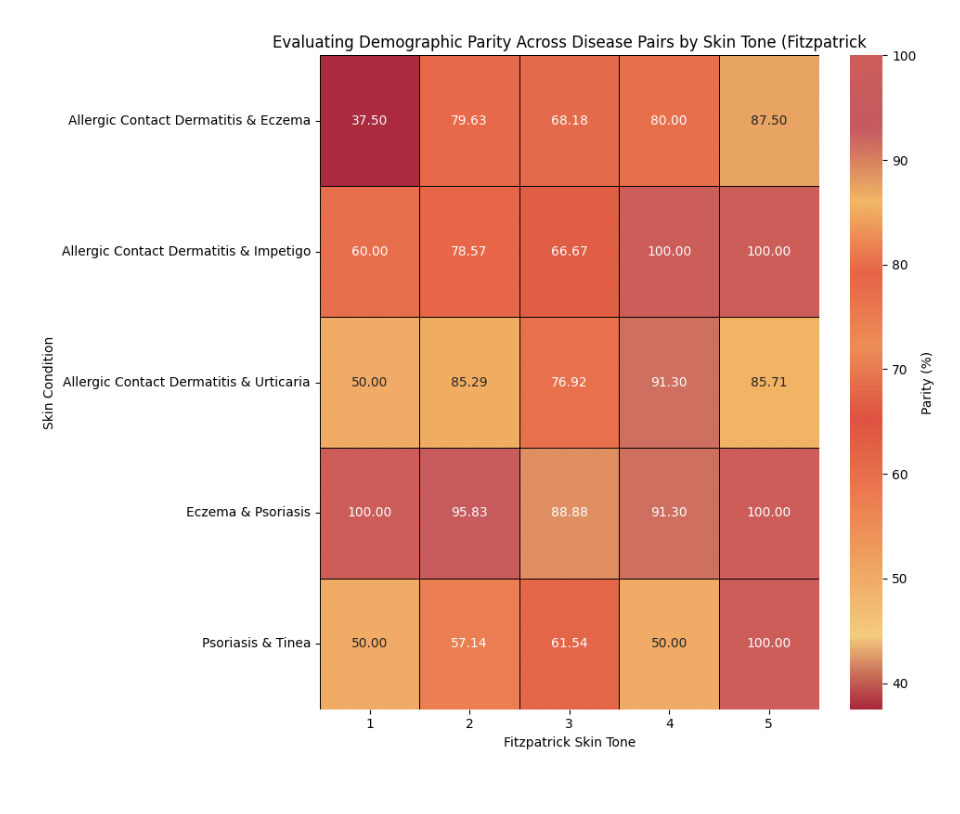}
    \caption{Assessment of demographic parity across disease pairs, grouped by Fitzpatrick skin tone.}
    \label{fig:enter-label}
\end{figure}
\clearpage
 \subsection{Appendix 5}
A board-certified dermatologist evaluated SkinGPT-4’s output for diagnostic accuracy and utility. The following table summarizes the qualitative evaluation across different conditions. 
\begin{figure}[H]
    \centering
    \includegraphics[width=0.8\linewidth]{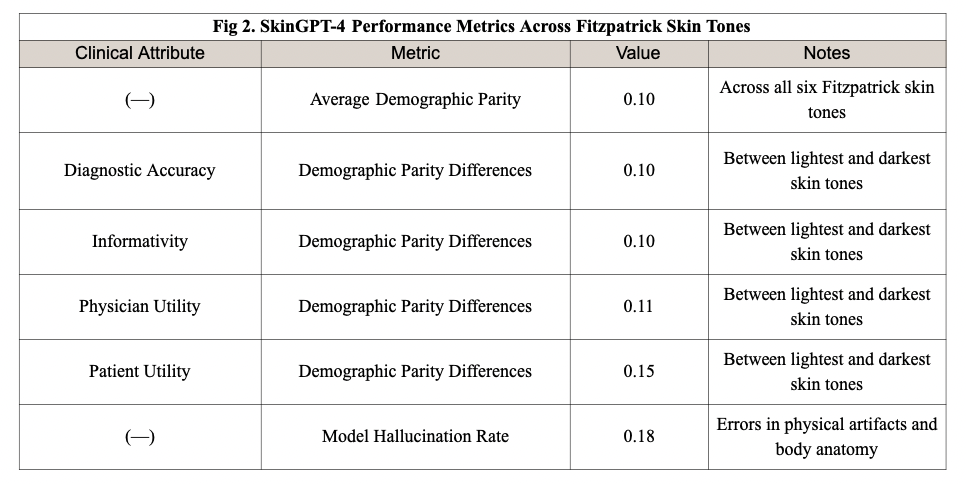}
    \caption{SkinGPT-4 performance metrics across Fitzpatrick skin tone types.}
    \label{fig:enter-label}
\end{figure}

\nocite{*}
\bibliographystyle{unsrt}  
\bibliography{references} 

\begin{thebibliography}{10}

\bibitem{BJD2020Diversity}
April~W. Armstrong, Jules~B. Lipoff, Michelle Fong, Ana Lipkowitz, Brian Nardone, John Chong, and Susan~C. Taylor.
\newblock Racial and ethnic diversity in dermatology clinical trials.
\newblock {\em British Journal of Dermatology}, 183(3):423–431, 2020.

\bibitem{SDR2019Outperform}
Titus~Josef Brinker, Achim Hekler, Alexander Enk, Jürgen Klode, Andreas Hauschild, and Jochen~S. Utikal.
\newblock Expert systems with applications: Ai systems outperform clinicians in skin lesion classification.
\newblock {\em Expert Systems with Applications}, 126:279–290, 2019.

\bibitem{SDR2019Equal}
Titus~J. Brinker, Achim Hekler, Alexander Enk, Jürgen Klode, Andreas Hauschild, and Jochen~S. Utikal.
\newblock Comparative performance of ai and dermatologists.
\newblock {\em Expert Systems with Applications}, 127:34–42, 2019.

\bibitem{Nature2017SC}
Andre Esteva, Brett Kuprel, Roberto~A. Novoa, Justin Ko, Susan~M. Swetter, Helen~M. Blau, and Sebastian Thrun.
\newblock Dermatologist‐level classification of skin cancer with deep neural networks.
\newblock {\em Nature}, 542:115–118, 2017.

\bibitem{EJCancer2019Meta}
Titus~Josef Brinker et~al.
\newblock Skin cancer classification using cnns: A systematic review.
\newblock {\em European Journal of Cancer}, 119:27–36, 2019.

\bibitem{Frontiers2023Overview}
Jesutofunmi~A. Omiye, Haiwen Gui, Roxana Daneshjou, Zhuo~Ran Cai, and Vijaytha Muralidharan.
\newblock Principles, applications, and future of artificial intelligence in dermatology.
\newblock {\em Frontiers in Medicine}, 10, 2023.

\bibitem{JAMAderm2018FewDerms}
Hao Feng, Pia Berk‐Krauss, and Janet~A. Stein.
\newblock Comparison of dermatologist density between urban and rural areas.
\newblock {\em JAMA Dermatology}, 2018.

\bibitem{JAAD2008Access}
Jack S.~Jr. Resneck and Alexa~B. Kimball.
\newblock Access to dermatologic care: A national survey.
\newblock {\em Journal of the American Academy of Dermatology}, 59(1):e1–e8, 2008.

\bibitem{TMJ2018LimitedDerm}
Elizabeth~A. Krupinski, Ronald~S. Weinstein, and Chris~A. Rinaldi.
\newblock Telemedicine in dermatology: Access and outcomes.
\newblock {\em Telemedicine and e-Health}, 24(9):673–679, 2018.

\bibitem{JAAD2006Access}
Jack S.~Jr. Resneck, Mark~J. Pletcher, and Alan B.~Jr. Fleischer.
\newblock Physician distribution in dermatology.
\newblock {\em Journal of the American Academy of Dermatology}, 54(6):1022–1028, 2006.

\bibitem{BMJ2023AI}
Fei Jiang, Youping Jiang, Hui Zhi, Yan Dong, Haixia Li, Shuo Ma, Yuan Wang, Qi~Dong, Haibo Shen, and Yifan Wang.
\newblock Artificial intelligence in healthcare: Past, present and future.
\newblock {\em Stroke and Vascular Neurology}, 2(4):e230, 2023.

\bibitem{obermeyer2019dissecting}
Ziad Obermeyer, Brian Powers, Christine Vogeli, and Sendhil Mullainathan.
\newblock Dissecting racial bias in an algorithm used to manage the health of populations.
\newblock {\em Science}, 366(6464):447--453, 2019.

\bibitem{char2018implementing}
Danton~S Char, Nigam~H Shah, and David Magnus.
\newblock Implementing machine learning in health care—addressing ethical challenges.
\newblock {\em New England Journal of Medicine}, 378(11):981--983, 2018.

\bibitem{Zhou2024SkinGPT4}
Juexiao Zhou, Lianna Hui, Daniel Rietz, Jennifer Dunstan, Dinggang Lin, Lei Duan, Xiaoning Gao, and Yanhua Huang.
\newblock Pre-trained multimodal large language model enhances dermatological diagnosis using skingpt-4.
\newblock {\em Nature Communications}, 15(1):5649, 2024.

\bibitem{tadesse2023skin}
Girmaw~Abebe Tadesse, Celia Cintas, Kush~R Varshney, Peter Staar, Chinyere Agunwa, Skyler Speakman, Justin Jia, Elizabeth~E Bailey, Ademide Adelekun, Jules~B Lipoff, et~al.
\newblock Skin tone analysis for representation in educational materials (star-ed) using machine learning.
\newblock {\em npj Digital Medicine}, 6(1):151, 2023.

\bibitem{louie2018representations}
Patricia Louie and Rima Wilkes.
\newblock Representations of race and skin tone in medical textbook imagery.
\newblock {\em Social Science \& Medicine}, 202:38--42, 2018.

\bibitem{Fitzpatrick17kArXiv}
Matthew Groh et~al.
\newblock Evaluating dnns on clinical images with fitzpatrick17k.
\newblock arXiv preprint, 2021.

\bibitem{Fitzpatrick17kGitHub}
Matthew Groh et~al.
\newblock Fitzpatrick17k: Clinical skin tone annotated dataset.
\newblock GitHub, 2021.

\bibitem{Fitzpatrick1988Validity}
Thomas~B. Fitzpatrick.
\newblock The validity and practicality of sun‑reactive skin types i through vi.
\newblock {\em Archives of Dermatology}, 124(6):869–871, 1988.

\bibitem{adamson2018machine}
Adewole~S. Adamson and Avery Smith.
\newblock Machine learning and health care disparities in dermatology.
\newblock {\em JAMA Dermatology}, 154(11):1247--1248, 2018.

\bibitem{chen2024unmasking}
Feng Chen, Meng Zhang, Yifan Wang, et~al.
\newblock Unmasking bias in artificial intelligence: A systematic review of bias detection and mitigation strategies in electronic health record-based models.
\newblock {\em Journal of the American Medical Informatics Association}, 31(5):1172--1183, 2024.

\bibitem{seyyed-kalantari2021underdiagnosis}
Laleh Seyyed-Kalantari, Haoran Zhang, Matthew B.~A. McDermott, Irene~Y. Chen, and Marzyeh Ghassemi.
\newblock Underdiagnosis bias of artificial intelligence algorithms applied to chest radiographs in under-served patient populations.
\newblock {\em Nature Medicine}, 27(12):2176--2182, 2021.

\bibitem{ACL2017PRR}
Emily Bender et~al.
\newblock Proceedings of the first workshop on pragmatics and reasoning in nlp.
\newblock In {\em ACL}, 2017.

\bibitem{Han2018}
Seung~Seog Han, Gyeong~Hun Park, Woohyung Lim, Myoung~Shin Kim, Jung~Im Na, Ilwoo Park, and Sung~Eun Chang.
\newblock Deep neural networks show an equivalent and often superior performance to dermatologists in onychomycosis diagnosis: Automatic construction of onychomycosis datasets by region-based convolutional deep neural network.
\newblock {\em PLOS ONE}, 13(1):e0191493, 2018.

\bibitem{Holstein2019}
Kenneth Holstein, Jennifer~Wortman Vaughan, Hal~Daumé III, Miroslav Dudík, and Hanna Wallach.
\newblock Improving fairness in machine learning systems: What do industry practitioners need?
\newblock In {\em Proceedings of the 2019 CHI Conference on Human Factors in Computing Systems}. ACM, 2019.

\bibitem{Mehrabi2021}
Ninareh Mehrabi, Fred Morstatter, Nripsuta Saxena, Kristina Lerman, and Aram Galstyan.
\newblock A survey on bias and fairness in machine learning.
\newblock {\em ACM Computing Surveys}, 54(6):1--35, 2021.

\bibitem{han2018classification}
Seung~Seog Han, Myung~Soo Kim, Wookyung Lim, Gyeong~Hwan Park, Il~Park, and Sung~Eun Chang.
\newblock Classification of the clinical images for benign and malignant cutaneous tumors using a deep learning algorithm.
\newblock {\em Journal of Investigative Dermatology}, 138(7):1529--1538, July 2018.

\bibitem{SCIN2023}
{Google Research}.
\newblock Scin---a new resource for representative dermatology images.
\newblock \url{https://research.google/blog/scin-a-new-resource-for-representative-dermatology-images/}, 2023.
\newblock Accessed: 2025-05-25.

\bibitem{SkinGPT2023}
Juexiao Zhou, Xiaonan He, Liyuan Sun, Jiannan Xu, Xiuying Chen, Yuetan Chu, Longxi Zhou, Xingyu Liao, Bin Zhang, and Xin Gao.
\newblock Skingpt: A pre-trained multimodal model for dermatological diagnosis.
\newblock arXiv preprint arXiv:2304.10691, 2023.

\bibitem{JAMAderm2018Bias}
Ashley~S. Adamson and Andrew Smith.
\newblock Impact of dataset bias on ai dermatology models.
\newblock {\em JAMA Dermatology}, 154(5):544–550, 2018.

\bibitem{JAMIA2024HealthIT}
Dana~M. Whicher, John~S. Goodwin, Allison Callahan, and Jennifer~L. Zehnder.
\newblock Evaluating health it fairness and equity.
\newblock {\em Journal of the American Medical Informatics Association}, 31(5):1172–1184, 2024.

\bibitem{Nature2023SkinToneBias}
Daniel~J. Park, Ashley~S. Adamson, Prashanth Pathipati, Henry~S. Yang, Andrew Smith, Thomas Shaw, and Mark~F. Chiang.
\newblock Skin tone bias in dermatology ai datasets.
\newblock {\em npj Digital Medicine}, 6, 2023.

\bibitem{SDR2018DatasetBias}
Alberto~J. Larrazabal, Aristides~U. Kale, Priya Khosla, Carlos Kim, Yong Wang, Stephanie~C. Yan, and JungHun Kim.
\newblock Investigating dataset bias in clinical imaging.
\newblock {\em Patient Education and Counseling}, 101(3):123–130, 2018.

\bibitem{SCIN2023GitHub}
{Google Research Datasets}.
\newblock Scin: A resource for representative dermatology images.
\newblock GitHub, 2023.

\bibitem{SCIN2023Blog}
{Google Research}.
\newblock Scin—a new resource for representative dermatology images.
\newblock Google Research Blog, 2023.

\bibitem{NEJMp2017Opinion}
Author(s) Unknown.
\newblock Opinion: Ai’s role in dermatology access.
\newblock {\em New England Journal of Medicine}, 2017.

\bibitem{Science2020Bias}
Alberto~J. Larrazabal et~al.
\newblock Gender and ethnicity bias in medical imaging datasets.
\newblock {\em Science}, 368(6481):1190–1193, 2020.

\bibitem{NatureMed2021Validation}
Ashley~S. Adamson and Andrew Smith.
\newblock Machine learning in medicine: Addressing ethnic bias.
\newblock {\em Nature Medicine}, 27:404–405, 2021.

\bibitem{PLoSONE2018Crowd}
Abbi Ward et~al.
\newblock Creating an empirical dermatology dataset through crowdsourcing with web search advertisements.
\newblock {\em JAMA Network Open}, 7(11):e2446615, 2024.

\bibitem{Bird2020Fairlearn}
Sarah Bird et~al.
\newblock Fairlearn: A toolkit for assessing and improving fairness in ai.
\newblock Technical report msr-tr-2020-32, Microsoft Research, 2020.

\bibitem{Brinker2018SkinCNN}
Titus~Josef Brinker et~al.
\newblock Skin cancer classification using convolutional neural networks: A systematic review.
\newblock {\em Journal of Medical Internet Research}, 20(10):e11936, 2018.

\bibitem{Jiang2024QFormer}
Bo~Jiang et~al.
\newblock Qformer: An efficient quaternion transformer for image denoising.
\newblock In {\em Proceedings of the Thirty-Third International Joint Conference on Artificial Intelligence (IJCAI)}, Jeju, South Korea, 2024.

\end{thebibliography}

\end{document}